\documentclass[twocolumn,english,aps,pra,showpacs]{revtex4}
\usepackage{amsmath, xcolor}
\usepackage{graphicx}
\usepackage{amssymb}
\usepackage{epsfig}

\usepackage{babel}
\usepackage{dsfont}

\renewcommand{\Re}{{\rm Re}}
\renewcommand{\Im}{{\rm Im}}
\newcommand{\Tr}{{\rm Tr}}
\newcommand{\rd}{{\rm d}}
\newcommand{\kb}{k_{\rm B}}

\newcommand{\be}{\begin{equation}}
\newcommand{\ee}{\end{equation}}
\newcommand{\bea}{\begin{eqnarray}}
\newcommand{\eea}{\end{eqnarray}}
\newcommand{\ba}{\begin{array}}
\newcommand{\ea}{\end{array}}

\newcommand{\rS}{{\rm S}}
\newcommand{\rP}{{\rm P}}

\newcommand{\rE}{{\rm E}}
\newcommand{\rH}{{\rm H}}

\newcommand{\rs}{{\rm s}}
\newcommand{\rp}{{\rm p}}

\newcommand{\re}{{\rm e}}
\newcommand{\ri}{{\rm i}}
\newcommand{\rt}{{\rm t}}

\begin{document}

\title{Radiative cooling of nanoparticles close to a surface }

\author{M. Tschikin}
\affiliation{Institut f\"{u}r Physik, Carl von Ossietzky Universit\"{a}t,
D-26111 Oldenburg, Germany.}

\author{S.-A. Biehs}
\affiliation{Institut f\"{u}r Physik, Carl von Ossietzky Universit\"{a}t,
D-26111 Oldenburg, Germany.}

\author{F.S.S. Rosa}
\affiliation{Laboratoire Charles Fabry,UMR 8501, Institut d'Optique, CNRS, Universit\'{e} Paris-Sud 11,
2, Avenue Augustin Fresnel, 91127 Palaiseau Cedex, France.}

\author{P. Ben-Abdallah}
\affiliation{Laboratoire Charles Fabry,UMR 8501, Institut d'Optique, CNRS, Universit\'{e} Paris-Sud 11,
2, Avenue Augustin Fresnel, 91127 Palaiseau Cedex, France.}

\date{\today}


\begin{abstract}
We study the radiative cooling of  polar and metallic nanoparticles immersed in a thermal bath
close to  a partially reflecting surface. The dynamics of relaxation is investigated at different 
distances from the surface, i.e., in the near-field and far-field zones. We demonstrate the 
existence of an oscillating behavior for the thermal relaxation time with respect to the separation 
distance from the surface, an analog of Friedel oscillations in Fermi liquids. 
\end{abstract}

\maketitle

\section{Introduction}  
It is well known since the pioneering works of Drexhage {\it et al.} \cite{DrexhageEtAl1968} 
and Chance {\it et al.}~\cite{ChanceEtAl1978} on the
molecular fluorescence emission that the radiative lifetime of excited atoms or molecules is not an intrinsic 
property of matter, but it depends on their close environment~\cite{Milonni, NovotnyHecht2006}. 
This de-excitation process is accompanied by spontaneous light emission, that can be investigated by either 
a classical or quantum approach. From a classical point of view, the atom/molecule is considered as a simple 
dipole that radiates as an antenna. The radiated field is then found with the help
of Maxwell's equations and the spontaneous decay rate is proportional to the partial local density of
states~\cite{FordWeber1984,Barnes1998}. On the other hand, from a quantum point of view, the process 
is described as a transition between different discrete states of the atom/molecule. The radiative lifetime or
its emission rate is then straightforwardly deduced by application of Fermi's 
golden rule~\cite{Milonni, NovotnyHecht2006,HenkelSandoghdar1998}.

The radiative cooling of a nano-object is, in a way, a continuous version of the
latter. When such an object is thermally excited at a given temperature, a continuum of modes
centered around the hot body's thermal frequency is excited according to the
Bose-Einstein distribution function. As a result, these modes de-excite throughout heat 
exchanges with the surrounding environment. As the local temperature decreases during the 
thermal relaxation process new modes are excited (at smaller frequencies), 
changing the cooling spectrum as the system is driven towards thermal equilibrium.
Understanding the underlying mechanisms for such a non-steady dissipative process
is of major importance for controling the heat flow in nanoscale systems and might
be of practical relevance in potential applications such as near-field 
thermo-photovoltaics~\cite{BasuEtAl2009} and thermal management in microelectronics~\cite{CareyEtAl2008}.

In this work, we describe the dynamic process of thermal relaxation for polar and metallic nanoparticles
facing a substrate at a given distance.  For the polar material we choose SiC because it supports 
surface phonon polaritons in the near-infrared. This is due to a negative real part of the permittivity 
for frequencies $\omega$ in the {\itshape reststrahlen} region, i.e. for $\omega_{\rm TO} < \omega < \omega_{\rm LO}$, 
where $\omega_{\rm TO}$ and $\omega_{\rm LO}$ are the frequencies of the transversal and longitudinal optical 
phonon, respectively. 
As we will discuss below the main channel for thermal relaxation will be due to these surface modes.
For the metal case we choose Au which can be described by a simple Drude model in the infrared region.
In particular, metals like Au have a negative real part of the permittivity for all frequencies $\omega$ smaller
than the plasma frequency $\omega_{\rm p}$ which is typically one or two orders of magnitudes larger than
$\omega_{\rm TO}$. In particular, for Au the surface plasmon polariton resonance lies in the ultraviolet region 
so that in this case the surface modes play a minor role for thermal relaxation. In fact, for metals the main
channel for thermal relaxation is due to the induction of eddy currents.

We compare the dynamic process for both materials considering only particles smaller than
the thermal wavelength $\lambda_{\rm th} = \hbar c /(\kb T)$, where $c$ is the vacuum velocity 
of light, $2 \pi \hbar$ is Planck's constant, $\kb$ is Boltzmann's constant, and $T$ is the
particle's temperature. Then we can use the dipole model based on Rytov's fluctuational electrodynamics~\cite{Rytovbook} 
in order to describe the heat loss
of the particle due to radiation. This model has for example been used to determine the
heat flux between a spherical nanoparticle and a flat surface~\cite{Dorofeyev1998,Pendry1999,MuletEtAl2001,VolokPersson2002,Martynenko2005,DedkovKyasov2007}, between a spherical particle and a structured or rough 
surface~\cite{BiehsEtAl2008,KittelEtAl2008,Rueting2010,BiehsGreffet2010}, between ellipsoidal particles and a flat or structured surfaces~\cite{HuthEtAl2010,BiehsHuthRueting2010} as well as between two or more spherical particles~\cite{ChapuisEtAl2008,PBA2006,PBA2008,PBA2011} or particles of arbitrary
shape~\cite{Domingues2005,Perez2008}. 
By using the dipole-model, we can introduce the thermal lifetime of a heated particle above a surface 
analogous to the lifetime of an excited atom or molecule above a surface. We investigate in particular the 
asymptotic behavior of cooling rates both in near- and far-field regimes and show how the cooling depends 
on material properties of the particle and the substrate. 

The paper is organized as follows: In section II we introduce the dipole model for the heat flux between a particle
and a flat surface. In section III we define the thermal relaxation time for small temperature gradients and discuss
it for different distance regimes for polar and metallic nanoparticles. Finally, in section IV we define
a general thermal relaxation time and solve the dynamical problem of thermal relaxation of a nanoparticle 
above a flat surface, numerically.


\section{Heat flux between a particle and a flat surface}
\begin{figure}[Hhbt]
  \epsfig{file=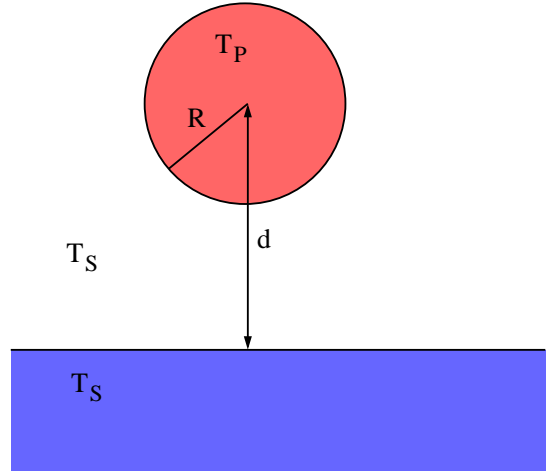, width = 0.4\textwidth}
  \caption{\label{Fig:Nanoparticle} Sketch of a nanoparticle with radius $R$ above a flat surface at a fixed distance $d$. The nanoparticle
           has the temperature $T_\rP$ which is assumed to be larger than the temperature of the surrounding $T_\rS$.}
\end{figure}

Let us consider the situation depicted in Fig.~\ref{Fig:Nanoparticle}. A spherical
nanoparticle with a radius $R$ is placed in front of a substrate at
a fixed distance $d$. Assuming isotropic, homogeneous and non-magnetic
materials, the nanoparticle and the substrate can be characterized
by the permittivities $\epsilon_{\rP}$ and $\epsilon_{\rS}$. Furthermore,
we assume that the particle's temperature ${T}_{\rP}$ is higher
than the temperature of its surroundings ${T}_{\rS}$.
Then the heat flux from the nanoparticle to the substrate due to the
imposed temperature gradient can be described within a dipole model~\cite{Dorofeyev1998,Pendry1999,MuletEtAl2001,VolokPersson2002} 
if the particle radius is much smaller than the thermal wavelength 
$\lambda_{{\rm th},\rP}$.
Including the magnetic response due to the induction of eddy currents in the object, 
which is important for metals in the infrared region~\cite{Martynenko2005,Tomchuk2006,DedkovKyasov2007,ChapuisEtAl2008}, the net power
dissipated by the dipole reads \cite{MuletEtAl2001}
\begin{equation}
  \mathcal{P}_{\rS \leftrightarrow \rP} = \sum_{i = \rE, \rH}\int_{0}^{\infty} \!\!\! \rd \omega\,2\omega\,\Im(\alpha^i) D^i (\omega,d) \triangle\Theta(\omega,T_{\rP},T_{\rS}),
\label{Eq:Power}
\end{equation}

\noindent where 
\begin{equation}
  D^i (\omega,d)=\frac{\omega}{\pi c^{2}} \Im \, \Tr \, \mathds{G}^i(\omega;d,d)
\label{Eq:LDOSDef}
\end{equation}
is the electric (i=E) and magnetic (i=H) local density of states (LDOS) at a distance $d$ from the surface~\cite{JoulainEtAl2003,Arvind2010}, and $\mathds{G}^i (\omega;z,z')$ is the dyadic Green tensor of the system~(see Appendix~\ref{App:GreensDyadic}). We also have
\begin{equation}
\begin{split}
  \triangle\Theta(\omega,T_\rP,T_\rS) 
       &= \Theta(\omega,T_{\rm P})-\Theta(\omega,T_{\rm S}) \\
       &= \frac{\hbar \omega}{e^{\hbar \omega / k_b T_\rP} -1} -  \frac{\hbar \omega}{e^{\hbar \omega / k_b T_\rS}-1} ,  
\end{split}
\end{equation}
which is the difference of the mean energy of harmonic oscillators at temperatures
${T}_\rP$ and ${T}_\rS$, respectively.  $\Im(\alpha^\rE)$ and $\Im(\alpha^\rH)$ are the imaginary parts of the 
electric and magnetic polarizability of the nanoparticle and will be specified in Eqs. (\ref{Eq:AlphaE})) and (\ref{Eq:AlphaM}). The physical interpretaion of 
expression (\ref{Eq:Power}) is the following one: The power radiated by a nanoemitter radiates 
into its environment is a function of the LDOS of the electromagnetic field at the 
emitter position, which defines all possible channels for radiative and non-radiative 
exchanges with the background. Note that the above expression does not take into 
account the interaction with the image dipole~\cite{VolokPersson2002}. This is reasonable
for distances $d^3 \gg R^3 |\epsilon_\rS - 1| |\epsilon_\rP - 1|/ (4 |\epsilon_\rP + 2| |\epsilon_\rS + 1|)$ - a
condition that is automatically fulfilled when $d \gg R$, where the dipole model is valid.

Substituting the known Green's functions for a halfspace geometry~\cite{ChenToTai,JoulainEtAl2005} into the definition of the LDOS in Eq.~(\ref{Eq:LDOSDef}) yields~\cite{JoulainEtAl2003}
\begin{align}
  D^\rE &= \frac{\omega^2}{2 \pi c^3} \int \!\! \frac{\rd^2 \kappa}{(2 \pi)^2} \, \frac{\re^{- 2 \Im(\gamma_0) d}}{|\gamma_0|^2} \biggl[
           \frac{\Re(\gamma_0)}{k_0}\biggl(  2 + \Re\bigl(r_\rs \re^{2 \ri \gamma_0 d}\bigr) \nonumber \\ 
        &\qquad+ \frac{2 \kappa^2 - k_0^2}{k_0^2} \Re\bigl(r_\rp \re^{2 \ri \gamma_0 d} \bigr) \biggr) + \frac{\Im(\gamma_0)}{k_0}\biggl( \Im(r_\rs) \nonumber \\
        &\qquad + \frac{2 \kappa^2 - k_0^2}{k_0^2} \Im(r_\rp) \biggr)\biggr]   \label{Eq:LDOSE}\\
  D^\rH &= \frac{\omega^2}{2 \pi c^3} \int \!\! \frac{\rd^2 \kappa}{(2 \pi)^2} \, \frac{\re^{- 2 \Im(\gamma_0) d}}{|\gamma_0|^2} \biggl[
           \frac{\Re(\gamma_0)}{k_0}\biggl(  2 + \Re\bigl(r_\rp \re^{2 \ri \gamma_0 d}\bigr) \nonumber \\ 
        &\qquad+ \frac{2 \kappa^2 - k_0^2}{k_0^2} \Re\bigl(r_\rs \re^{2 \ri \gamma_0 d} \bigr) \biggr) + \frac{\Im(\gamma_0)}{k_0}\biggl( \Im(r_\rp) \nonumber \\
        &\qquad + \frac{2 \kappa^2 - k_0^2}{k_0^2} \Im(r_\rs) \biggr)\biggr] \label{Eq:LDOSH} 
\end{align}
where we have introduced the usual Fresnel coefficients for s- and p-polarized waves~\cite{Jackson}
\begin{align}
  r_\rs &=  \frac{\gamma_0 - \gamma_\rS}{\gamma_0 + \gamma_\rS},  \label{Eq:rp}\\
  r_\rp &=  \frac{\gamma_0 \epsilon_\rS - \gamma_\rS}{\gamma_0 \epsilon_\rS + \gamma_\rS} \label{Eq:rs}
\end{align}
with the z-component of the wave vector in vacuum $\gamma_0 = \sqrt{k_0^2 - \kappa^2}$, 
the z-component of the wave vector inside the substrate $\gamma_\rS = \sqrt{k_0^2 \epsilon_\rS - \kappa^2}$, and the wave number in vacuum $k_0 = \omega/c$.
Note, that the electric LDOS can be retrieved from the magnetic LDOS by interchanging $r_\rs \leftrightarrow r_\rp$ and vice versa. Furthermore, the electric and magnetic LDOS contain the contribution of propagating modes for lateral wave vectors $\kappa < k_0$ for which $\Im(\gamma_0) = 0$ and the contribution of evanescent modes for which  $\kappa > k_0$ and therefore $\Re(\gamma_0) = 0$.
Due to the rotational invariance with respect to rotations around the $z$-axis, we can reduce (\ref{Eq:LDOSE}) and 
(\ref{Eq:LDOSH}) to one dimensional integrals by introducing cylindrical coordinates.

Finally, we need the polarizabilities of the nanoparticle to evaluate expression~(\ref{Eq:Power}).
In the case of a spherical particle of radius~$R$ the polarizabilities can be derived from 
Mie scattering theory~\cite{BohrHuff}. Denoting the particle's relative permittivity by $\epsilon_{\rm P}$ 
and introducing the dimensionless variables $x = k_0 R$ and $y = \sqrt{\epsilon_\rP} k_0 R$, 
one finds~\cite{ChapuisEtAl2008,BohrHuff} 
\begin{align}
	\alpha_{\rm P}^{\rm E} & = 
	2\pi R^3 \frac{\left(2 \epsilon_{\rm P} + 1\right)\left[\sin(y) - y \cos(y)\right]
	                -y^2\sin(y)}
		       {\left(\epsilon_{\rm P} - 1\right)\left[\sin(y) - y \cos(y)\right]
	                +y^2\sin(y)} \label{Eq:AlphaE}\\
    	\alpha_{\rm P}^{\rm H} & = 
	\frac{\pi R^3}{3}\left(\frac{\left(x^2-6\right)}{y^2}\!\!\left[y^2 + 3 y \cot(y) - 3\right] 
	   - \frac{2 x^2}{5}\right) 
        \label{Eq:AlphaM}
\end{align}
for $x \ll 1$, implying that the particle's radius should be small compared
to the dominant thermal wavelength, $R \ll \lambda_{\rm th}$. If we demand 
further that $|y| \ll 1$, i.e. that the radius should be smaller than the skin depth
at thermal frequencies, the above expressions reduce to
\begin{align}
  	\alpha_{\rm P}^{\rm E} & = 
	4\pi\,R^3 \frac{\epsilon_{\rm P}-1}{\epsilon_{\rm P}+2} \; , 
\label{Eq:Clausius-Mossotti}\\
  	\alpha_{\rm P}^{\rm H} & = 
	\frac{2\pi}{15} R^3\left(k_0 R\right)^2
	\left(\epsilon_{\rm P}-1\right) \; .
  \label{Eq:Eddy_currrents}
\end{align}
We have checked that with the previous expressions we retrieve the dipole contribution to the heat flux 
for an isolated nanoparticle, i.e., $r_\rP = r_\rS = 0$, derived by means 
of Rytov's fluctuational electrodynamics ~\cite{KattawarEisner1970}. 
Note that in the whole model we have for convenience neglected any 
nonlocal effects both in the substrate~\cite{FordWeber1984,JoulainHenkel2006,Dorofeyev2006,ChapuisEtAl2008nonloc,HaakhHenkel2011} 
and the particle~\cite{Ruppin1975,Ruppin1992}.


\section{Thermal relaxation time of a small particle}  

In the absence of phase transitions, the time evolution of the temperature field of a
nanoparticle of heat capacity ${C_\rp}$, mass density ${\rho}$ and of volume $V = 4 \pi R^3 / 3$ is
given by
\begin{equation}
  \rho C_{\rp} V \frac{\rd T_{\rP}}{\rd t} = -\mathcal{P}_{\rS \leftrightarrow \rP}{(T_{\rP})}.
\label{Eq:Energy}
\end{equation} 
which is just a heat-balance equation. It is important to point out that
the substitution of (\ref{Eq:Power}) in the r.h.s of (\ref{Eq:Energy}) is apparently
inconsistent, as we are using an expression obtained 
from the fluctuation dissipation theorem - and therefore assuming local thermal
equilibrium - to calculate the cooling rate of a given object, which is clearly a
process out-of-equilibrium (even locally). Further investigation, however, shows
that here this does not cause any trouble: the characteristic time $\tau_{\rm ph}$
for temperature homogenization throughout the body ($\tau_{\rm ph} \lesssim 10^{-8} s$) 
is, as we shall see, much smaller than the thermal relaxation
time (to be defined below). That essentially means that the particle cools down in a
quasi-stationary fashion, which in turn justifies the use of (\ref{Eq:Power}) into
(\ref{Eq:Energy}).   

Considering a slightly heated nanoparticle with a temperature  $T_\rP = T_\rS + \Delta T$, 
a thermal relaxation time $\tau_{\ell}$  (i.e., the inverse of cooling rate) can 
be defined by linearization of (\ref{Eq:Energy}) with respect to the temperature difference
$\Delta T$.  The resulting linearized equation
\begin{equation}
  \frac{\rd T_{\rP}}{\rd t} =- \frac{1}{\tau_{\ell}} \Delta T
\label{Eq:Linearized_Energy}
\end{equation} 
 together with (\ref{Eq:Power}) leads to
\begin{equation}
\tau_{\ell}^{-1} = \frac{1}{\rho C_{p} V} \sum_{i = \rE, \rH}\intop_{0}^{\infty} \rd\omega\, 2 \omega \Im(\alpha^i)D^i(\omega,z)   \left. \frac{\rd \Theta(\omega,T)}{\rd T}\right|_{T_\rS}.
\label{Eq:Lifetime1}
\end{equation} 

\begin{figure}[Hhbt]
  \epsfig{file = 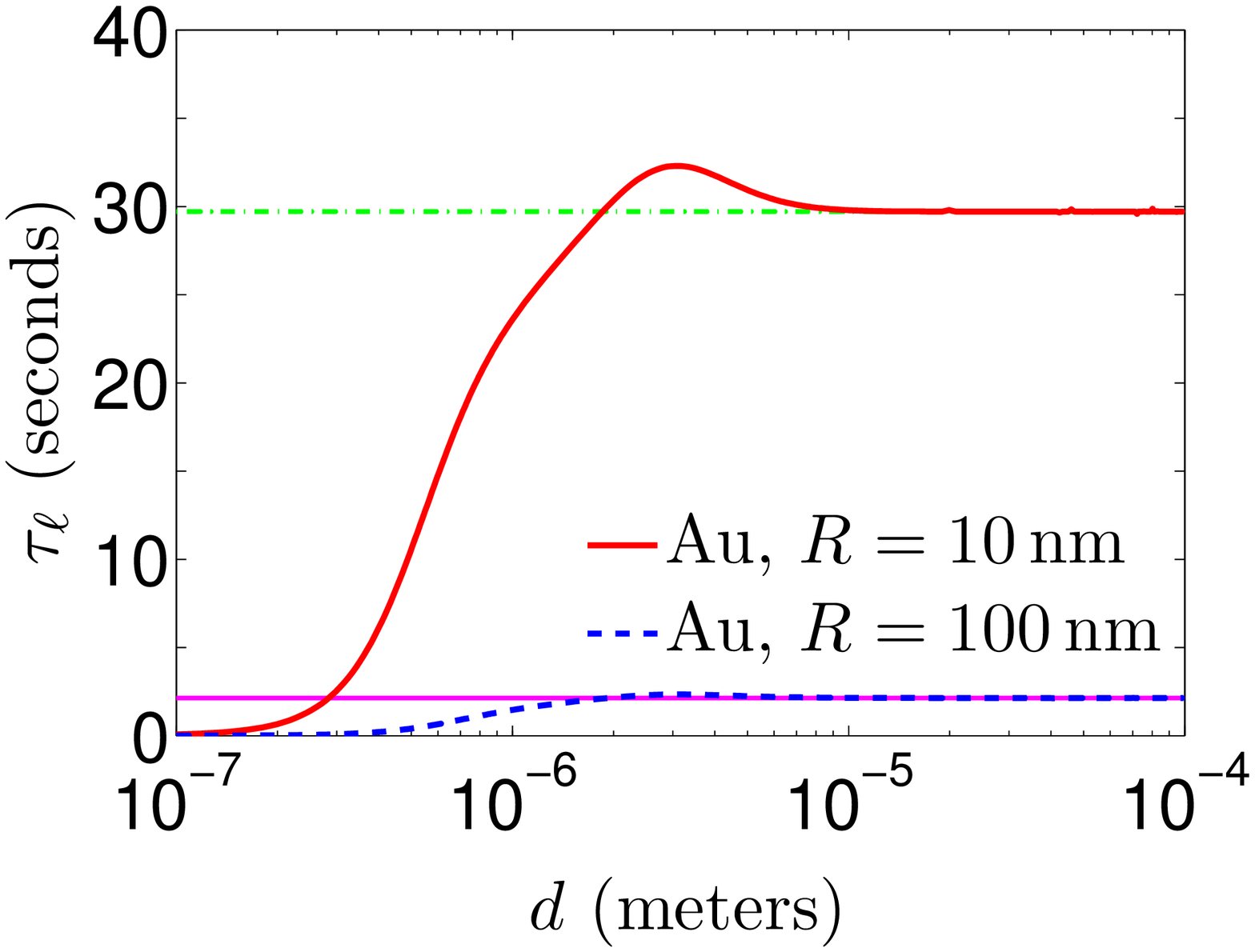, width = 0.39\textwidth}
  \epsfig{file = 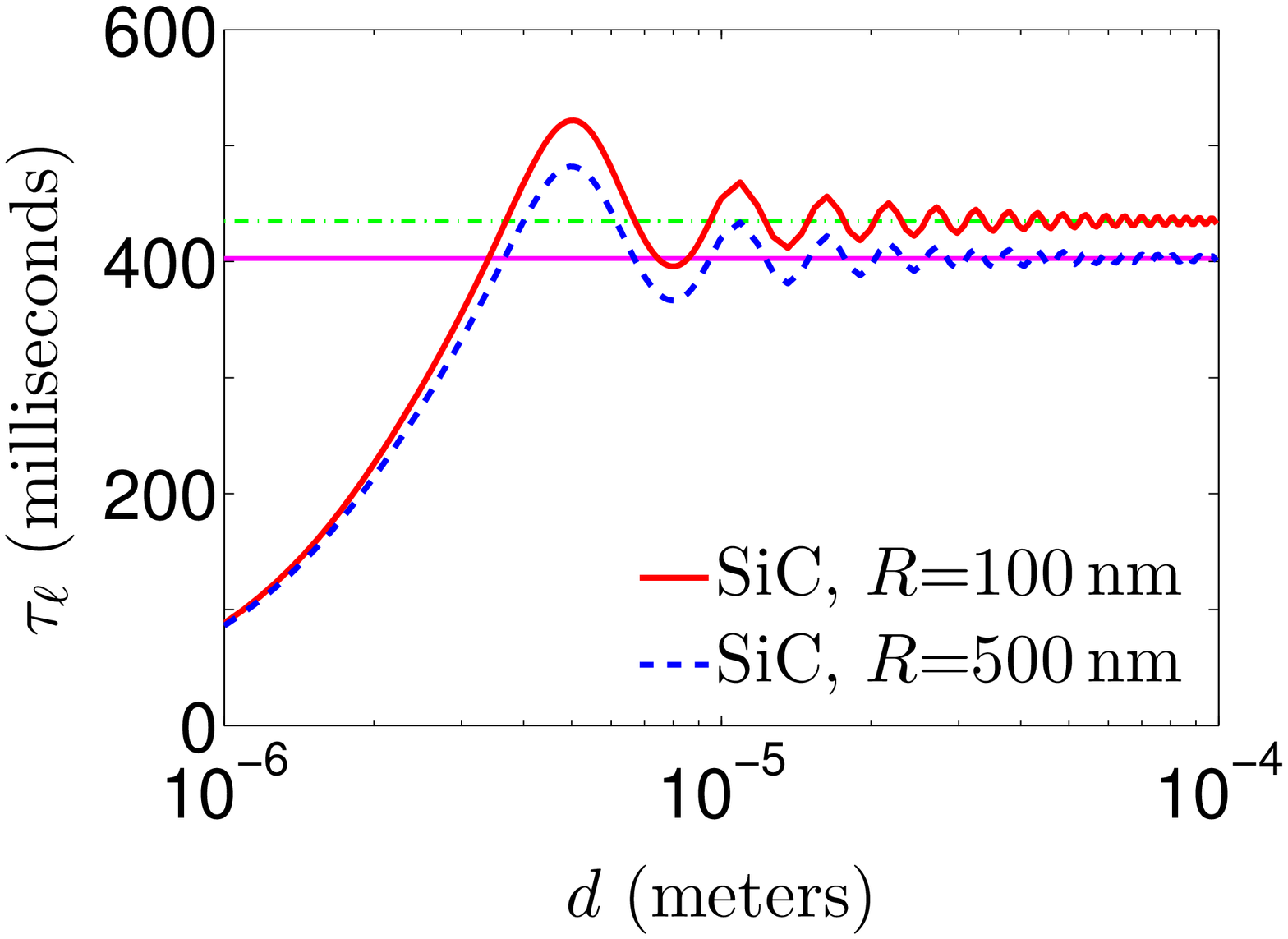, width = 0.39\textwidth}
  \caption{\label{Fig:Lifetime} Semi-logarithmic plot of the distance dependence of the TRT of a nanoparticle 
           above a substrate  with temperature $T_\rS = 300\,{\rm K}$
           (a) for a gold nanoparticle above a gold surface, (b) a SiC nanoparticle above a SiC surface. 
            The horizontal lines mark the asymptotic values given by Eq. (\ref{Eq:Lifetime2}).
	   For $\epsilon_\rP/\epsilon_\rS$ we
           use for Au the Drude model and for SiC a Drude-Lorentz model~\cite{BohrHuff,ShchegrovEtAl2000}. For $\rho$ we use 
           $\rho^{\rm Au}$ and $\rho^{SiC}$ and for the heat capacity we use a simple 
           Debye model (see appendix~\ref{App:HeatCapacity})
           so that we have at $ T_\rP = 300\,{\rm K}$ the values 
           $\rho^{\rm Au} C_\rp^{\rm Au} = 2.404\cdot10^6 \,{\rm J} {\rm m}^{-3} {\rm K}^{-1}$ and 
           $\rho^{\rm SiC} C_\rp^{\rm SiC} = 2.212\cdot10^6 \,{\rm J} {\rm m}^{-3} {\rm K}^{-1}$.}
\end{figure}

In Fig.~\ref{Fig:Lifetime} we show the evolution of the  thermal relaxation time
(TRT) with respect to the separation distance $d$ from the surface for different materials 
obtained using Eq.~(\ref{Eq:Lifetime1})
together with Eqs.~(\ref{Eq:LDOSE})-(\ref{Eq:AlphaM}). 
We first note the presence of two distinct regions for the TRT. When the separation
distance is much smaller than the thermal wavelength $\lambda_{{\rm th}, \rP}$ we see that the
cooling process is faster than for an isolated particle and it decreases drastically 
as the separation distance is reduced. On the other hand, 
in the intermediate regime for distances $d \approx \lambda_{{\rm th}, \rP}$
we see that the TRT oscillates with respect to $d$ and approaches  for large $d$ the (linearized) value $\tau_{0, \ell}$ of an isolated particle, given by~\cite{KattawarEisner1970,BohrHuff}
 \begin{equation}
\tau_{0, \ell}^{-1}= \frac{1}{\rho C_{p} V}\sum_{i = \rE, \rH}\intop_{0}^{\infty}\rd\omega\, \left(\frac{\omega^{3}}{\pi^{2}c^{3}} \right) \Im(\alpha^i)  \left. \frac{\rd \Theta(\omega,T)}{\rd T}\right|_{T_\rS} .
\label{Eq:Lifetime2}
\end{equation} 
This behavior of the cooling rate demonstrates that we can either reduce or increase the power exchange
with the surface for $d \approx \lambda_{{\rm th}, \rP}$. In order to get more physical insight, we calculate below the asymptotic  behavior of the power exchanged between the substrate and the nanoparticle in these two regimes.

\subsection{Cooling in the near-field regime}  

The asymptotic expressions for the LDOS in the quasistatic regime, i.e., for distances $z \ll \lambda_{{\rm th}, \rP}$ can be derived by considering only contributions for wave vectors $\kappa \gg k_0$. Then one gets the
known expressions~\cite{JoulainEtAl2003} 
\begin{align}
  D^\rE (\omega,d) &\approx \frac{\omega^2}{\pi^2 c^3} \frac{\Im(\epsilon_\rS)}{|\epsilon_\rS + 1|^2} \frac{1}{4 (k_0 d)^3}, \\
  D^\rH (\omega,d) &\approx \frac{\omega^2}{\pi^2 c^3} \biggl[ \frac{\Im(\epsilon_\rS)}{4} + \frac{\Im(\epsilon_\rS)}{|\epsilon_\rS + 1|^2} \biggr] \frac{1}{4 (k_0 d)}. 
\end{align}
and therefore
\bea
\label{TRTNF}
\hspace{-0.5 cm} \tau_{\ell}^{-1} \hspace{-8pt}&&\sim \frac{1}{\rho C_{p} V}  \frac{1}{4 (k_0 d)} \intop_{0}^{\infty} d\omega \frac{2 \omega^3}{\pi^2 c^3} \Im(\epsilon_\rS)  \left. \frac{\rd \Theta(\omega,T)}{\rd T}\right|_{T_\rS} \times \nonumber \\ 
&&\hspace{-10pt}\Bigg\{ \frac{1}{(k_0 d)^2}\frac{\Im(\alpha^E)}{|\epsilon_\rS + 1|^2} + \Im(\alpha^H)\left[ \frac{1}{4}+ \frac{1}{|\epsilon_\rS + 1|^2} \right] \Bigg\}.
\eea
From the structure of the previous expression we see clearly that the main contributions to $1/\tau_\ell$
come from the particle's electric and magnetic resonances [which are located at $\epsilon_\rP\approx-2$ and 
$\epsilon_\rP \rightarrow \infty$  as can be seen in Eqs. (\ref{Eq:Clausius-Mossotti})-(\ref{Eq:Eddy_currrents})] 
and from the evanescent modes in the surface
(at $\epsilon_\rS\approx-1$). In other words, the decrease in the
TRT can be attributed to the enhanced radiative heat transfer due to the coupling 
of the particle's (induced) dipole moment to the evanescent modes of the surface~\cite{MuletEtAl2001,ChalopinEtAl2011}. In addition,
it is also seen from (\ref{TRTNF}) that, depending on the strengths of the electric 
and magnetic polarizabilities, the lifetime can be proportional to $d^3$ or $d$.  For metals we typically
observe the latter, as the induction of Foucault's currents gives rise to a big magnetic contribution to the heat transfer~\cite{Tomchuk2006,ChapuisEtAl2008,HuthEtAl2010}, whereas for polar materials like SiC the electric LDOS dominates the heat flux~\cite{ShchegrovEtAl2000,MuletEtAl2001}, producing a $1/d^3$ behavior.


\subsection{Cooling in the intermediate and far-field regimes}  

On the other hand, for distances $d \gtrsim \lambda_{{\rm th}, \rP}$ we can derive the 
appropriate expressions by using the stationary phase method~(see appendix~\ref{App:SatPhas}).
We then obtain
\begin{align}
  D^\rE (\omega,d)\approx \frac{\omega^2}{\pi^2 c^3} \biggl[ \frac{1}{2} + \Im\biggl( \frac{1 - \sqrt{\epsilon_\rS}}{1 + \sqrt{\epsilon_\rS}} \frac{\re^{2 \ri k_0 d}}{4 k_0 d} \biggr) \biggr], \label{Eq:StatPhas1}\\
  D^\rH (\omega,d)\approx \frac{\omega^2}{\pi^2 c^3} \biggl[ \frac{1}{2} - \Im\biggl( \frac{1 - \sqrt{\epsilon_\rS}}{1 + \sqrt{\epsilon_\rS}} \frac{\re^{2 \ri k_0 d}}{4 k_0 d} \biggr) \biggr].\label{Eq:StatPhas2}
\end{align}
We see that the presence of the substrate makes the electric and magnetic LDOS oscillate around 
one half of the free space value, given by $\omega^2 / 2 \pi^2 c^3$. 
This behavior is analogous to the fermionic density oscillation close to a defect in a Fermi fluid or 
to the electric charge screening  in a pool of ions~\cite{JoulainEtAl2003}. From (\ref{Eq:StatPhas1})-(\ref{Eq:StatPhas2}) we get a simplified expression
expression for the TRT, namely
\begin{equation}
\label{TRTFF}
\begin{split}
\tau_{\ell}^{-1} &\approx \frac{1}{\rho C_{p} V} \intop_{0}^{\infty} \rd\omega\,
           2 \biggl( \frac{\omega^3}{\pi^2 c^3} \biggr)  \left. \frac{\rd \Theta(\omega,T)}{\rd T}\right|_{T_\rS} \\
           &\quad\times \biggl\{ \Im(\alpha^\rE) \biggl[ \frac{1}{2} + \Im\biggl( \frac{1 - \sqrt{\epsilon_\rS}}{1 + \sqrt{\epsilon_\rS}} \frac{\re^{2 \ri k_0 d}}{4 k_0 d} \bigr) \biggr] \\
          &\qquad + \Im(\alpha^\rH) \biggl[ \frac{1}{2} - \Im\biggl( \frac{1 - \sqrt{\epsilon_\rS}}{1 + \sqrt{\epsilon_\rS}} \frac{\re^{2 \ri k_0 d}}{4 k_0 d} \bigr) \biggr] \biggr\}.
\end{split}
\end{equation}
In the case where we have narrow dipolar resonances we can write 
\be
\label{Approx}
\intop_{0}^{\infty} \rd\omega\, \Im(\alpha^{\rE,\rH}) f(\omega) \approx \sum_{i=1}^{N_{\rE,\rH}} f(\omega_i^{\rE,\rH}) \intop_{0}^{\infty} \rd\omega\, \Im(\alpha^{\rE,\rH}),
\ee
where  $f(\omega)$ is a smooth and arbitrary function that varies slowly on the scale of the widths of the resonances of $\Im(\alpha^{\rE,\rH})$. Here $\omega_i^{\rE}$ ($\omega_i^{\rH}$) are the electric (magnetic) dipole 
resonances of the particle and
$N_{\rE}$ ($N_{\rH}$) is their total number, and therefore we can simplify $\tau_\ell^{-1}$
even further,  
\begin{eqnarray}
\label{TRTFF2}
\tau_{\ell}^{-1} &&\hspace{-9pt}\approx  F(\omega^{\rE}, T_\rS) \left[ \frac{1}{2} + \Im\biggl( \frac{1 - \sqrt{\epsilon_\rS}}{1 + \sqrt{\epsilon_\rS}} \frac{\re^{2 \ri k_{\rE} d}}{4 k_{\rE} d} \bigr) \right]  \intop_{0}^{\infty} \rd\omega\, \Im(\alpha^{\rE})\nonumber  \\
&&\hspace{-31pt}+\,  F(\omega^{\rH}, T_\rS) \left[\frac{1}{2} - \Im\biggl( \frac{1 - \sqrt{\epsilon_\rS}}{1 + \sqrt{\epsilon_\rS}} \frac{\re^{2 \ri k_{\rH} d}}{4 k_{\rH} d} \bigr) \right]  \intop_{0}^{\infty} \rd\omega\, \Im(\alpha^{\rH}) ,
\end{eqnarray}
where 
\be
\label{Def}
F(\omega, T_\rS) = \frac{1}{\rho C_{p} V} \biggl( \frac{2\omega^3}{\pi^2 c^3} \biggr)
 \left. \frac{\rd \Theta(\omega,T)}{\rd T}\right|_{T_\rS} ,
\ee
$k_{\rE} = \omega^{\rE}/c$, $k_{\rH} = \omega^{\rH}/c$, and we assumed that only one resonance of each
type (E and H) is present. Expression (\ref{TRTFF2}) then shows clearly the behavior of the TRT
for large distances (and narrow resonances): it oscillates in a superposition of two periods while
decaying as $1/d$. In particular, when one polarizability strongly dominates the other 
[say, $F(\omega^{\rE},  T_\rS) \gg F(\omega^{\rH},  T_\rS)$], we should see a clean ``monochromatic" sinc-like oscillation of period $2\pi/k_{\rE}$.
Since this is precisely the case for SiC, that is exactly what we observe in Figure 2b.
That we don't have a similar behavior for gold is explained by (i) the fact that for Au particles
we have $\omega^{\rE} \gg \omega_{\rm th}$ for $ T_\rS \approx 300 K$, meaning that the electric dipole
gives no contribution to the heat flux (because $F(\omega^{\rE},300) \approx 0$),
and (ii) the very large width of the magnetic resonance, which invalidates approximation
(\ref{Approx}) and in practice smears out the magnetic contribution to the TRT, resulting in the
behavior seen in Figure 2a.

\subsection{Temperature dependence}

\begin{figure}[Hhbt]
  \epsfig{file = 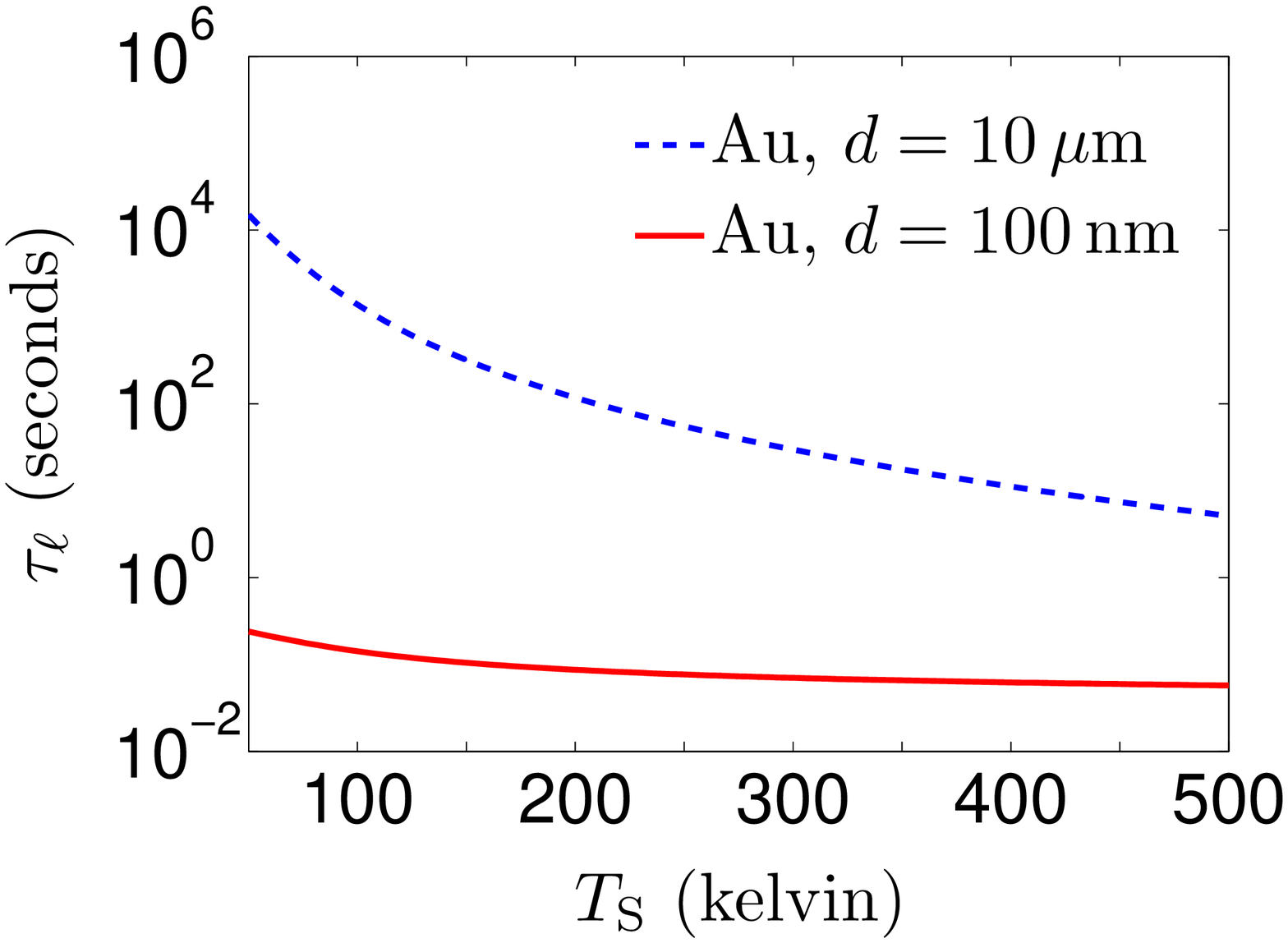, width = 0.4\textwidth}
  \epsfig{file = 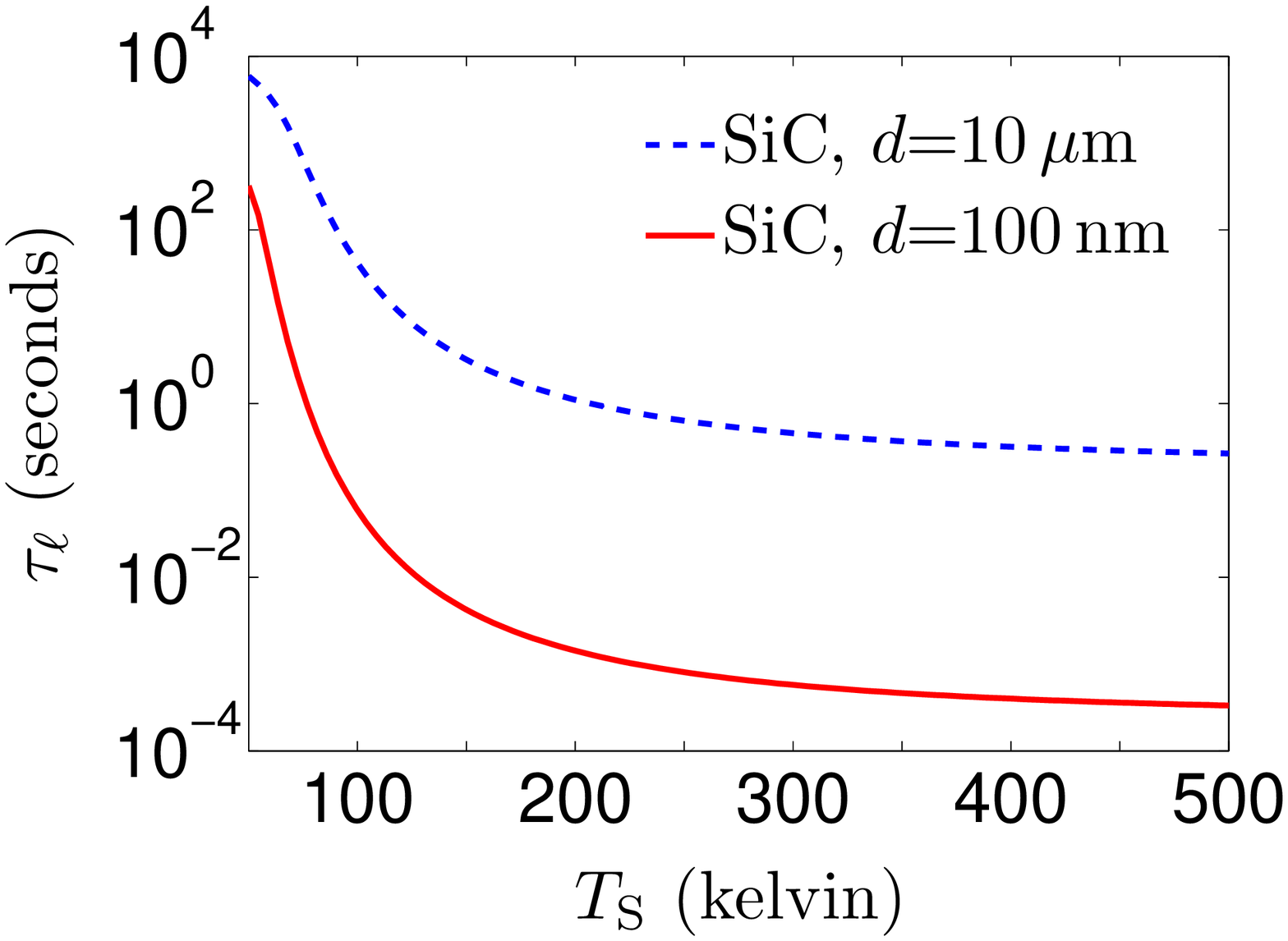, width = 0.4\textwidth}
  \caption{\label{Fig:TempDep} Semi-logarithmic plot of the temperature dependence of the TRT of a nanoparticle above a substrate at
           a fixed distance $d = 10\, \mu{\rm m}$ and $d = 100\,{\rm nm}$ for 
           (a) a gold nanoparticle above a gold surface, (b) a SiC nanoparticle above a SiC surface. The particle 
           radius is in both cases $R = 10\,{\rm nm}$.}
\end{figure}

Finally, we study the temperature dependence of the TRT. In Fig.~\ref{Fig:TempDep} we show some
numerical results of the TRT over the temperature for $d = 10\, \mu{\rm m}$ and $d = 100\,{\rm nm}$, i.e., 
in the far-field and near-field regimes neglecting the temperature dependence of the material
parameters~\cite{Dorofeyev2008}. First of all, one can observe that the TRT decreases with
increasing temperatures $T_\rS$. Hence, the hotter the environment, 
the faster the particle cools down (reminding that, in the linear approximation, 
$T_\rP = T_\rS + \Delta T$ with $\Delta T \ll T_\rS$). Note that in
the given temperature range of $50-500\,{\rm K}$ the TRT varies enormously. 
For Au (SiC) particles at $d = 10\,\mu{\rm m}$ it can be as big as $4$ hours ($98$ minutes)
at $T_\rS=50 K$, dropping all the way down to around $5$ seconds ($266$ milliseconds) at $T_\rS=500 K$. 
Even more interesting is what happens in the near-field regime ($d = 100\,{\rm nm}$), where the TRT
variation is not so impressive for gold (from about $240$ milliseconds at $T_\rS = 50 K$ to $58$ milliseconds at 
$T_\rS = 500 K$) but still very large for SiC (around $5$ minutes at $T_\rS = 50 K$ and $0.3$ milliseconds 
at $T_\rS = 500 K$). This difference at close distances is readily understood in terms of the coupling
to the surface modes of the substrate. For SiC we have surface waves that lie in the frequency region
close to the maximum of the thermal function $F(\omega,300)$ defined in (\ref{Def}),
and so the heat exchange is quite large at those temperatures.
By decreasing $ T_\rS$, however, we shift $F(\omega, T_\rS)$
toward lower frequencies and strongly suppress the participation of the surface modes in the heat transfer,
leading to the huge increase in the TRT. By contrast, the surface modes for Au lie at much higher frequencies,
which means that they don't play any significant role in the heat flux in the considered range and therefore
can't affect much the TRT.

\section{Dynamics of relaxation}   

The TRT defined in Eq.~(\ref{Eq:Lifetime1}) is only useful for small temperature
differences $\Delta T$, since it is based on a linearization procedure.
Here, we consider now the full dynamical process regarding the cooling of a particle by
solving numerically the nonlinear Eq.~(\ref{Eq:Energy}). Here,
we need to define the TRT in the general case: Let us assume that the nanoparticle is heated 
up with respect to its surrounding ($T_\rP > T_\rS$), then we define the thermal relaxation 
time of the particle as the elapsed time $\tau$ the particle needs to cool down 
to $T_\rS + \Delta T/\rm{e}$. This definition ensures that for very small temperature differences we retrieve the linearized relaxation time $\tau_{\ell}$ defined in Eq.~(\ref{Eq:Lifetime1}).


\subsection{Cooling dynamics for an isolated particle} 

\begin{figure}[Hhbt]
  \epsfig{file = 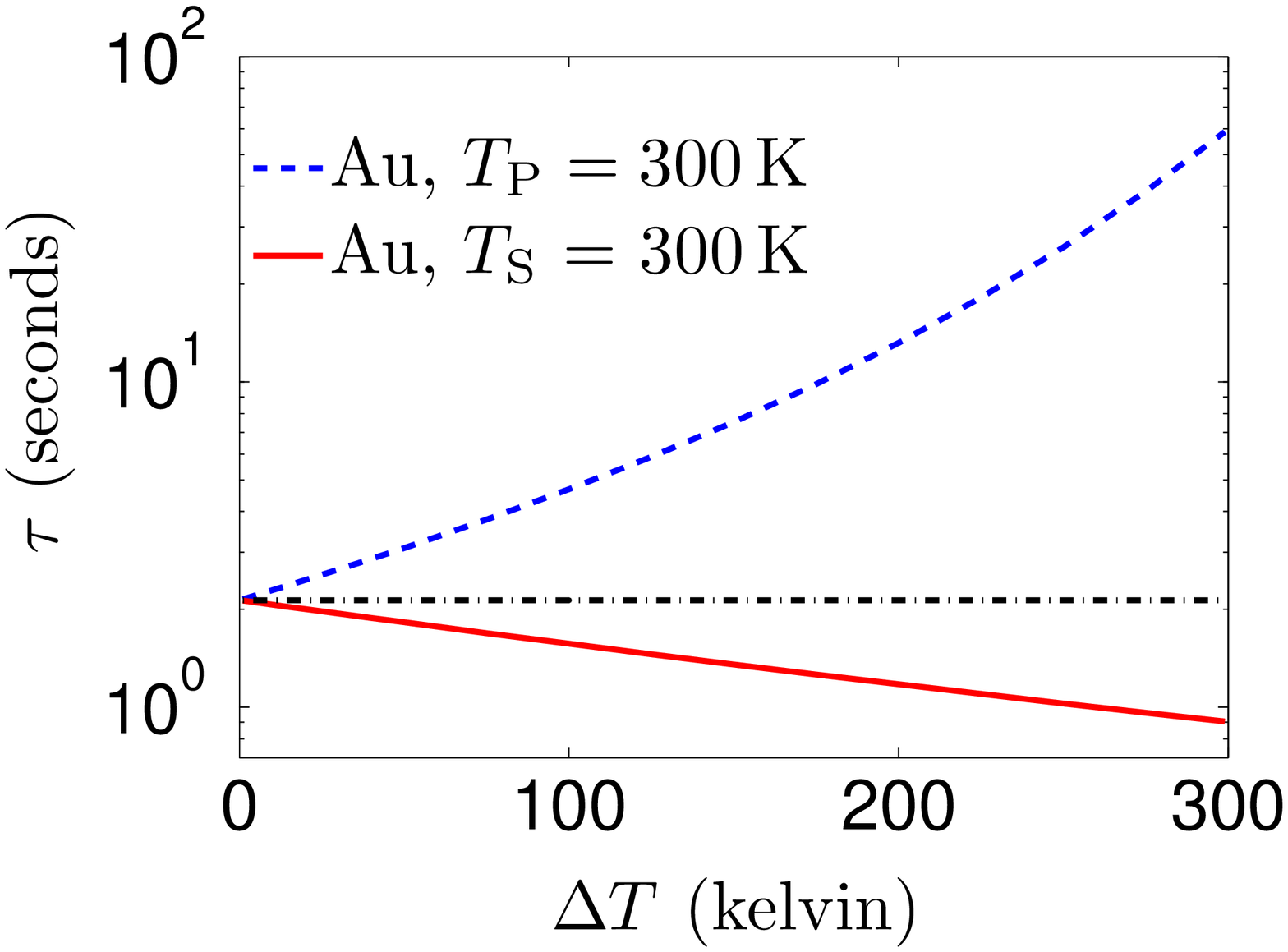, width = 0.4\textwidth}
 \epsfig{file = 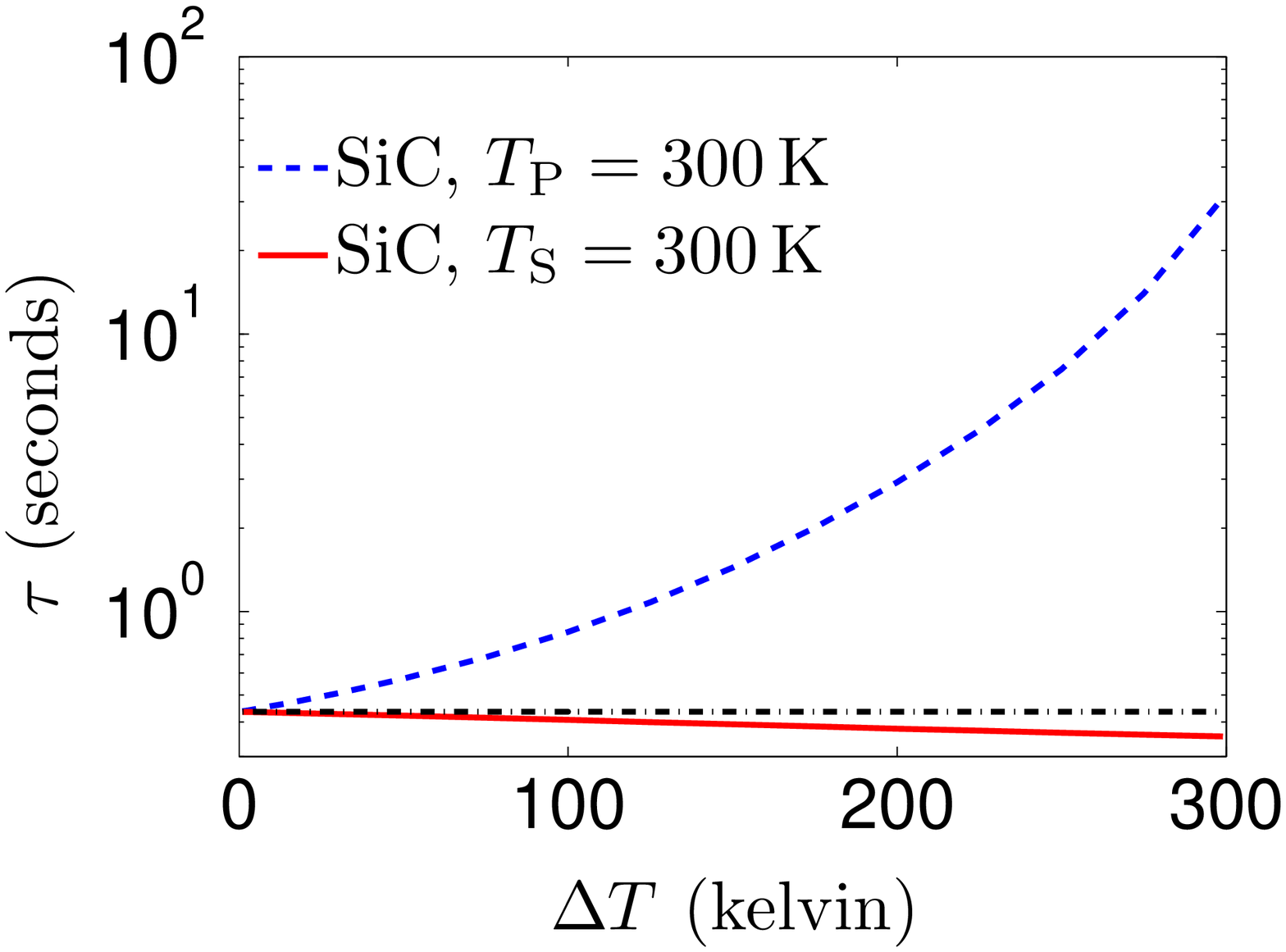, width = 0.4\textwidth}
  \caption{\label{Fig:Partisole}  Semi-logarithmic plot of the TRT of a gold (a) and a SiC (b) nanoparticle with radius $R = 100\,{\rm nm}$ in free space, as a function of the temperature difference $\Delta T = T_\rP(0) - T_\rS$ (see text for further details). The horizontal line is the corresponding linearized TRT $\tau_{\ell}$.  }
\end{figure}

For an isolated particle we already saw that $D^\rE = D^\rH = D_0/2$,
so that the cooling dynamics is governed by the following equation
\begin{equation}
  \frac{\rd T_\rP}{ \rd t}=- \frac{1}{\rho C_{p}V} \sum_{i = \rE,\rH}\intop_{0}^{\infty}\rd\omega\, \biggl( \frac{\omega^{3}}{\pi^{2}c^{3}} \biggr) \Im(\alpha^i)\triangle\Theta(\omega,T_{\rP},T_{\rS})\\.
\label{Eq:isole}
\end{equation}
In Fig.~\ref{Fig:Partisole} we show the results obtained by a numerical integration
of Eq.~(\ref{Eq:isole}) using a Runge-Kutta method with adaptive time steps. As for the initial conditions, 
we consider two different cases: (i) The solid lines show the TRT (as defined above) for $T_\rS = 300 K$ 
and $T_\rP(t=0) = 300 K + \Delta T$, while (ii) the dotted lines correspond to $T_\rS = 300 K - \Delta T$ 
and $T_\rP(t=0) = 300 K$. It can be seen in Fig.~\ref{Fig:Partisole} that in the first case the TRT decreases with respect to $\Delta T$, whereas in the second case it increases. This somewhat unexpected result
shows that the cooling rate strongly  depends not only on the temperature {\it difference} between particle and environment, but also on the {\it absolute} temperature of the environment. The first scenario can be 
qualitatively understood from the fact that by heating up the particle (while $T_\rS$ is kept constant) the radiated power increases nonlinearly with $T_\rP$, leading therefore to a faster cooling. In the second case the radiative power of the particle at $t = 0$ is kept constant while $T_\rS$ is cooled down. Thus the increasing difference between $T_\rP$ and $T_\rS$ leads to an increase of the time the particle needs to cool down to $T_\rS + \Delta T/\rm{e}$.

\subsection{Cooling dynamics of a dipole near a surface} 

\begin{figure}[Hhbt]
  \epsfig{file = 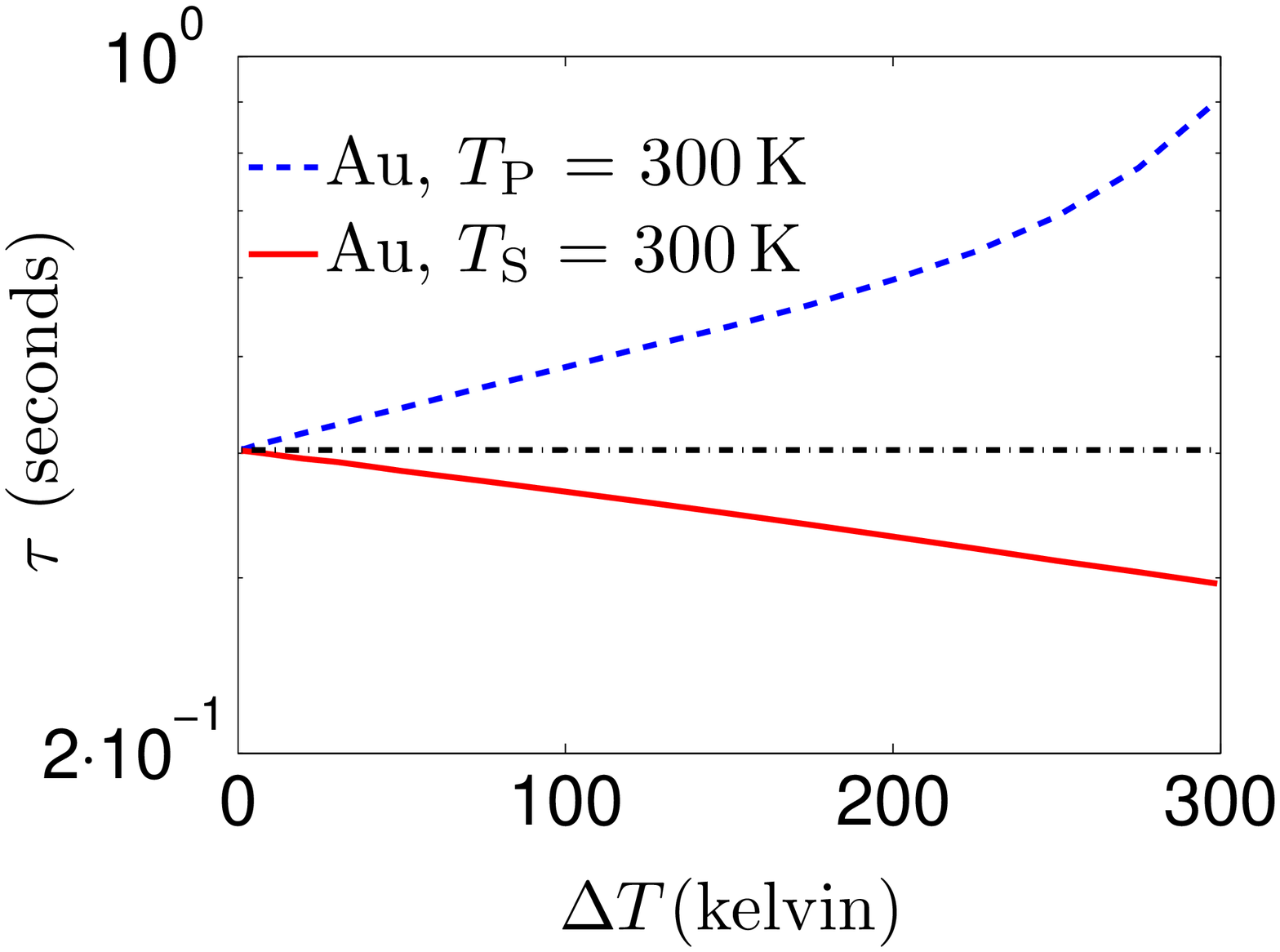, width = 0.4\textwidth}
  \epsfig{file = 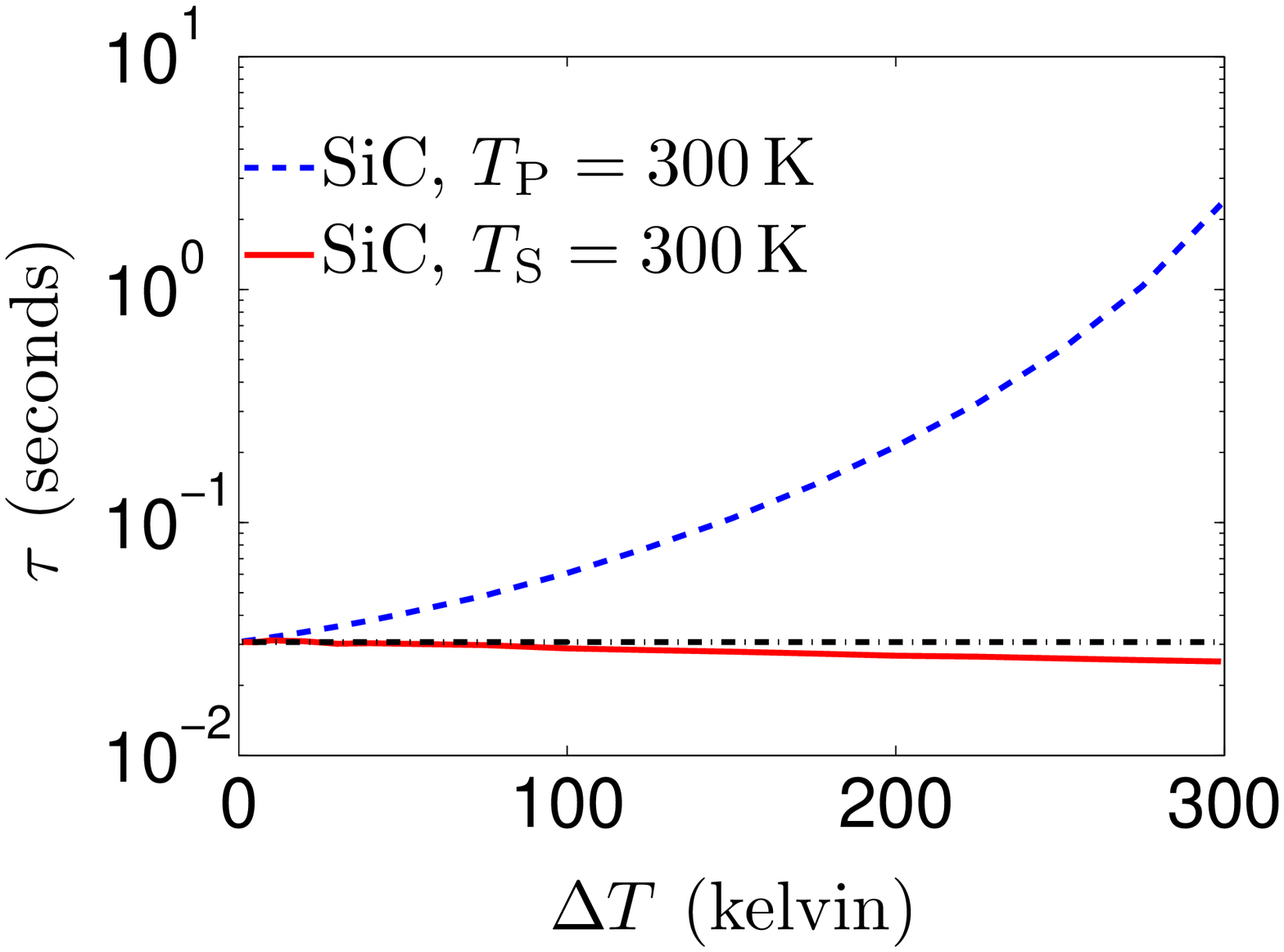, width = 0.4\textwidth}
  \caption{\label{Fig:Partsurface} As in Fig.~\ref{Fig:Partisole} but for a Au (SiC) particle above a Au (SiC) surface, at a distance of $d = 500\,{\rm nm}$. }
\end{figure}

Now we look at the evolution of the temperature of a slightly heated nanoparticle near a
surface. In Fig.~\ref{Fig:Partsurface} we show the time evolution of $T_\rP$ for a gold nanoparticle 
with $R = 100\,{\rm nm}$ for a distance $d=500\,$nm above a gold surface and a SiC nanoparticle above a SiC surface.
As previously we consider a fixed particle temperature or a fixed surrounding temperature. 
As for the isolated particle, the exact relaxation times are in very good agreement with their linearized counterparts for
$\Delta T \rightarrow 0$. 
We observe that due to the presence of the interface the TRT is much smaller than for the isolated particles.
Apart from this we find the same qualitative behaviour as in the isolated case in Fig.~\ref{Fig:Partisole}.

\section{Conclusions}

We have investigated the thermal relaxation of a nanoparticle close to an interface. In particular, we have
introduced a relaxation time $\tau_\ell$ for the regime of small temperature differences between the particle and
the surface and a generalized relaxation time $\tau$ for arbitrary temperature differences. Equipped 
with these definitions we have studied numerically the thermal relaxation of gold and SiC nanoparticles close to a gold or SiC surface, respectively.  We have shown that the relaxation time behaves similarly to the lifetime of an atom close to a surface, varying rapidly for distances smaller than the thermal wavelength and finally droping to zero. In addition, we found that the thermal relaxation time $\tau_\ell$ is very sensitive to the temperature and can vary drastically - by several orders of magnitude - in the temperature range
of $50 - 500\,{\rm K}$. We have identified all mechanisms which govern the thermal relaxation from subwavelength separation distances (in near-field regime) all the way to long separation distances (in far field regime). 
 
This work can be generalized to the heating/cooling of a nano-object immersed in a more complex environment and exploited for the thermal management at nanoscale of complex plasmonic systems. Finally, a time-resolved scanning near-field optical microscopy as the one used in Refs.~\cite{RousseauEtAl2009,vZwol2011} could be used to measure the time evolution of a substrate temperature in non-equilibrium situation by detecting, at the scale of TRT, the time variation of diffracted signal.


\begin{acknowledgements}
 M.\ T.\ gratefully acknowledges support from the Stif\-tung der Metallindustrie im Nord-Westen. 
       P.\ B.-A.\ and F.S.S.\ R.\ acknowledges the support of the Agence Nationale de la Recherche through the Source-TPV
       project ANR 2010 BLANC 0928 01.
\end{acknowledgements}

\appendix

\section{Green's dyadic for a flat interface}
\label{App:GreensDyadic}

The Green's dyadic function with observation point $\mathbf{r} = (x,y,z)^t$ and 
source point $\mathbf{r}' = (x',y',z')^t$ above the flat surface, i.e.,
for $0 < z \leq z'$ can be expressed as~\cite{ChenToTai,Sipe1987,JoulainEtAl2005}
\begin{equation}
\begin{split}
  \mathds{G}^{(0)}(\mathbf{r,r'};\omega) &= \int \!\!\!\frac{\rd^2 \kappa}{(2 \pi)^2} \frac{\ri \re^{\ri \boldsymbol{\kappa\cdot(\mathbf{x - x'})}}}{2 \gamma_0} \\
                                         &\qquad\times\biggl[ \mathds{1}_{--} \re^{\ri \gamma_0 (z' - z)} 
                                          + \mathds{R}_{+-} \re^{\ri \gamma_0 (z' + z)} \biggr] \\
                                          &\quad - \frac{1}{3 k_0^2} \delta(z - z') \delta(\mathbf{x - x'}) \mathbf{e}_z \otimes \mathbf{e}_z  
\end{split}
\label{Eq:AppGreenFlat}
\end{equation} 
where $ \mathbf{e}_z$ is the unit vector in $z$-direction and $\otimes$ symbolizes the dyadic product. The
tensors $\mathds{1}$ and $\mathds{R}$ are defined as
\begin{align}
  \mathds{1}_{--} &= \sum_{j = \{\rs,\rp\}} \hat{\mathbf{a}}_j^-(\boldsymbol{\kappa}) \otimes \hat{\mathbf{a}}_j^-(\boldsymbol{\kappa}) \\
  \mathds{R}_{+-} &=  \sum_{j = \{\rs,\rp\}} r_j \hat{\mathbf{a}}_j^+(\boldsymbol{\kappa}) \otimes \hat{\mathbf{a}}_j^-(\boldsymbol{\kappa})
\end{align}
where 
\begin{align}
  \hat{\mathbf{a}}_\rs^-(\boldsymbol{\kappa}) &= \hat{\mathbf{a}}_\rs^+ (\boldsymbol{\kappa}) = \frac{1}{\kappa} (-k_y, k_x,0)^\rt , \\
  \hat{\mathbf{a}}_\rp^-(\boldsymbol{\kappa}) &= - \frac{1}{\kappa k_0} (k_x \gamma_0, k_y \gamma_0, \kappa^2)^\rt , \\
  \hat{\mathbf{a}}_\rp^+(\boldsymbol{\kappa}) &= \frac{1}{\kappa k_0} (k_x \gamma_0, k_y \gamma_0, -\kappa^2)^\rt
\end{align}
are the polarization vectors for s- and p-polarization. Note that these vectors are always orthogonal, but only have a unitary norm for
propagating modes with $\kappa < k_0$. The reflection coefficients $r_\rs$ and $r_\rp$ are the usual Fresnel coefficients
in Eqs.~(\ref{Eq:rp}) and (\ref{Eq:rs}).

\section{Density of states in the intermediate and far-field regimes}
\label{App:SatPhas}

We describe here the asymptotic expansion of the TRT in the far-field regime.
To this end we start from the expressions for the electric and magnetic LDOS
from Eqs.~(\ref{Eq:LDOSE}) and (\ref{Eq:LDOSH}) and take only the propagating modes into account.
The distance independent part can be separated from the distance dependent part, 
so that the LDOS can be written as
\begin{align}
  D^\rE(\omega) &= \frac{D_0}{2} + \Re \int_0^{k_0} \frac{\rd^2\kappa}{(2 \pi)} \,
                          \frac{D_0}{4 k_0 \gamma_0}\biggl[ r_\rs + \frac{2 \kappa^2 - k_0^2}{k_0^2} r_\rp \biggr] \re^{2 \ri \gamma_0 d} \\
  D^\rH(\omega) &= \frac{D_0}{2} + \Re \int_0^{k_0} \frac{\rd^2\kappa}{(2 \pi)} \,
                          \frac{D_0}{4 k_0 \gamma_0}\biggl[ r_\rp + \frac{2 \kappa^2 - k_0^2}{k_0^2} r_\rs \biggr] \re^{2 \ri \gamma_0 d}
\end{align}
where $D_0 = \omega^2/(\pi^2 c^3)$. That means for the distance dependent part 
we have two-dimensional integrals of the form ($i = \rE, \rH$)
\begin{equation}
  I_i  = \Re \int\rd^2 \kappa\, F_i(\kappa) \re^{\ri \phi(\kappa)}
\end{equation}
which have a highly oscillating integrand for $\phi(\kappa)$ larger than one, 
which corresponds in our case to distances larger than the wavelength.
By using the stationary phase method~\cite{Cho} such integrals can be approximated by
\begin{equation}
  I_i \approx \Re \frac{2 \pi}{\ri \sqrt{a c - b^2}} F_i(\kappa_\rs) \re^{\ri \phi(\kappa_\rs)}
\label{Eq:StatPhas}
\end{equation}
where $\kappa_\rs$ is defined such that
\begin{equation}
  \frac{\partial \phi}{\partial k_x} \bigg|_{\kappa_\rs} = \frac{\partial \phi}{\partial k_y} \bigg|_{\kappa_\rs} = 0
\end{equation}
and
\begin{equation}
  a = \frac{\partial^2 \phi}{\partial k_x^2} \bigg|_{\kappa_\rs}, \quad 
  b = \frac{\partial^2 \phi}{\partial k_x \partial k_y} \bigg|_{\kappa_\rs},\, \text{and} \quad
  c = \frac{\partial^2 \phi}{\partial k_y^2} \bigg|_{\kappa_\rs}.
\end{equation}
In our case $\phi(\kappa) = 2 \gamma_0 d$ so that $\boldsymbol{\kappa}_\rs = (0,0)^\rt$ and $a = c = 2 d/k_0$ and $b = 0$.
Hence, by inserting these results into Eq.~(\ref{Eq:StatPhas}) we find the far-field expressions in
Eqs.~(\ref{Eq:StatPhas1}) and (\ref{Eq:StatPhas2}).

\section{Heat Capacity}
\label{App:HeatCapacity}

In order to describe the heat capacity of the two considered materials (Au and SiC)
we use a standard Debye model~\cite{Ashcroft}
\begin{equation}
  C_{\rm V} = 9 n_{\rm a} \frac{ N_{\rm A} \kb}{M} \biggl( \frac{T}{\Theta_{\rm D}} \biggr) \int_0^{\Theta_{\rm D}/T} \rd x\, 
              \frac{x^4 \re^{x}}{(\re^x - 1)^2},
\end{equation}
where $N_{\rm A} = 6.022 \cdot 10^{23}\, {\rm mol}^{-1}$ is Avogadro's constant, $M$ is the molar mass and
$n_{\rm a}$ the number of atoms or ions per unit cell, and $\Theta_{\rm D}$ is Debye's temperature. For convenience
we assume that $C_{\rm V} \approx C_{\rm p}$. When using the parameters $\Theta_{\rm D}^{\rm Au} = 170\,{\rm K}$,
$ n_{\rm a}^{\rm Au} = 1$, $M^{\rm Au} = 196.97\,{\rm g} \, {\rm mol}^{-1}$ and 
 $\Theta_{\rm D}^{\rm SiC} = 1100\,{\rm K}$, $ n_{\rm a}^{\rm SiC} = 2$, $M^{\rm SiC} = 40.1\,{\rm g}\, {\rm mol}^{-1}$ 
we get values for $C_{\rm p}$ which are in very good agreement with experimental and numerical 
results~\cite{GeballeGiauque1952,KarchEtAl1994,ZywietzEtAl1996,PorterYip1997}.




\end{document}